\documentclass[showpacs,prc,preprint,nofootinbib,showkeys]{revtex4}
\pdfoutput=1

\usepackage{bm}
\usepackage{amsmath}
\usepackage{graphicx}
\usepackage{subfigure}
\usepackage[colorlinks=true,linktocpage=true,linkcolor=blue,citecolor=blue]{hyperref}
\usepackage[usenames]{color}

\newcommand{\bel}[1]{\begin{eqnarray}\label{#1}}
\newcommand{\eel}{\end{eqnarray}}
\newcommand{\rf}[1]{Eq.~(\ref{#1})}
\newcommand{\rfm}[1]{Eqs.~(\ref{#1})}
\newcommand{\rfn}[1]{(\ref{#1})}

\def\Lq{\Lambda_q}

\def\Lg{\Lambda_g}

\def\xq{\xi_q}
\def\xg{\xi_g}

\def\Lq{\Lambda_Q}

\def\Lg{\Lambda_G}

\def\xq{\xi_Q}
\def\xg{\xi_G}

\newcommand{\s}{{\rm s}}		
\newcommand{\Qp}{{Q^+}}		
\newcommand{\Qm}{{Q^-}}		
\newcommand{\G}{{G}}		
\newcommand{\Qpm}{{Q^\pm}}		
\newcommand{\Q}{{Q}}		
\newcommand{\eq}{{\rm eq}}  
\newcommand{\an}{{\rm a}}  
\newcommand{\teq}{\tau_\eq}

\newcommand{\bpT}{{\boldsymbol p}_T} 
 
\newcommand{\I}{{\cal I}}

\newcommand{\pT}{p_{T}}
\newcommand{\pL}{p_{L}}

\newcommand{\lp}{\left(}
\newcommand{\rp}{\right)}

\newcommand{\lsb}{\left[}
\newcommand{\rsb}{\right]}

\newcommand{\bea}{\begin{eqnarray}}
\newcommand{\eea}{\end{eqnarray}}
\newcommand{\beal}[1]{\begin{eqnarray}\label{#1}}
\newcommand{\eeal}{\end{eqnarray}}
\newcommand{\nn}{\nonumber}
\newcommand{\f}[2]{\frac{#1}{#2}}
\newcommand{\EQ}[1]{Eq.~(\ref{#1})}

\newcommand{\EQSTWO}[2]{Eqs.~(\ref{#1})~and~(\ref{#2})}

\newcommand{\p}{\partial}
\newcommand{\twpt}{(\tau,w,\pT)}

\newcommand{\VP}{\vphantom{\frac{}{}}\!}

\begin{document}
 

\title{Anisotropic-hydrodynamics approach to \\ a quark-gluon fluid mixture}

\author{Wojciech Florkowski} 
\affiliation{Institute of Nuclear Physics, Polish Academy of Sciences, PL-31342 Krak\'ow, Poland}
\affiliation{Institute of Physics, Jan Kochanowski University, PL-25406~Kielce, Poland} 

\author{Ewa Maksymiuk} 
\affiliation{Institute of Physics, Jan Kochanowski University, PL-25406~Kielce, Poland} 

\author{Radoslaw Ryblewski} 
\affiliation{Institute of Nuclear Physics, Polish Academy of Sciences, PL-31342 Krak\'ow, Poland}

\date{\today}


\begin{abstract}
Anisotropic-hydrodynamics framework is used to describe a mixture of quark and gluon fluids.
The effects of quantum statistics, finite quark mass, and finite baryon number density are taken into account.
The results of anisotropic hydrodynamics are compared with exact solutions of the coupled kinetic equations
for quarks and gluons in the relaxation time approximation. The overall very good agreement between
the hydrodynamic and kinetic-theory results is found.
\end{abstract}

\pacs{25.75.-q,  12.38.Mh, 25.75.Ld, 24.10.Nz, 47.75.+f,}

\keywords{relativistic heavy-ion collisions, quark-gluon plasma, Boltzmann equation, relativistic hydrodynamics, anisotropic hydrodynamics}

\maketitle 

\section{Introduction}
\label{sec:i}

In this work we generalise several earlier results concerning application of anisotropic hydrodynamics (aHydro) to model the behaviour of a quark-gluon fluid mixture. The first study of this type was restricted to massless particles obeying classical statistics  and  based on the kinetic equations in the relaxation time approximation (RTA)~\cite{Florkowski:2012as}. Further developments included the use of a modified RTA collision term \cite{Florkowski:2013uqa} as well as comparisons with the exact solutions of the coupled kinetic equations \cite{Florkowski:2014txa}. The latter work revealed problems connected with the exclusive use of the zeroth and first moments of the kinetic equations to construct the framework of aHydro. These problems were solved in  \cite{Florkowski:2015cba} by relaxing some of the previous assumptions  restricting the number of hydrodynamic parameters and by using, in addition, the second moments of the kinetic equations, as suggested in \cite{Tinti:2013vba}.

In the meantime, comparisons between predictions of various formulations of dissipative hydrodynamics and exact solutions of the kinetic theory have become a popular and convenient tool to check the validity range of the hydrodynamic frameworks~\cite{Florkowski:2013lza,Florkowski:2013lya,Bazow:2013ifa,Florkowski:2014sfa,Florkowski:2014sda,Denicol:2014tha,Denicol:2014xca,Nopoush:2014qba,Florkowski:2015lra,Molnar:2016gwq,Martinez:2017ibh,Damodaran:2017ior}. The latter are now commonly used to interpret the experimental data collected in heavy-ion collisions at RHIC and the LHC~\cite{Romatschke:2009im,Ollitrault:2012cm,Gale:2013da,Jeon:2015dfa,Jaiswal:2016hex}. Comparisons between hydrodynamic-model predictions and kinetic-theory results allow us to analyse relations between effective hydrodynamic models and microscopic underlying theories~\cite{Noronha:2015jia,Heller:2016rtz,Denicol:2016bjh,Bemfica:2017wps}, for a recent review see~\cite{Florkowski:2017olj}. 

In this work we use our recent results on the exact solutions of the coupled quark-gluon kinetic equations~\cite{Florkowski:2017jnz}. In general, the new solutions found in~\cite{Florkowski:2017jnz} describe massive quarks obeying Fermi-Dirac statistics and coupled to massless gluons obeying Bose-Einstein statistics. Moreover, Ref.~\cite{Florkowski:2017jnz} includes consistently the effects of finite baryon number density. In order to reproduce some of the earlier results, the cases where classical statistics is used and/or where quarks are treated as massless particles have been also analysed in~\cite{Florkowski:2017jnz}. The present study of aHydro refers mainly to the general case of massive quarks and quantum statistics.

Similarly to earlier works on mixtures we restrict ourselves to boost invariant systems~\cite{Bjorken:1982qr}, since this assumption allows for finding exact solutions of the kinetic-theory equations. For the sake of simplicity, we also assume that the relaxation time is constant and the same for quarks and gluons. The use of a temperature dependent relaxation time, which is a natural choice for conformal theories, increases substantially the computational time for the exact calculations but, otherwise, does not introduce an important restriction for the performed comparisons. 

Our results confirm that aHydro is a very good approximation for the kinetic-theory results provided an appropriate set of the moments of kinetic equations is selected  to construct the aHydro framework --- this includes using the second moments and special combinations of the zeroth moments, as in Refs.~\cite{Florkowski:2015cba} and \cite{Tinti:2013vba}. We find that starting from different initial conditions aHydro reproduces very well the time dependence of several physical quantities, in particular, of the ratio of the longitudinal and transverse pressures. This brings further arguments in favour of using the framework of aHydro for phenomenological modeling of heavy-ion collisions~\cite{Alqahtani:2017jwl}.

\bigskip
In this paper we use $x^\mu = (t,x,y,z)$ and  $p^\mu = (p^0 = E_p, p_x, p_y, p_z=\pL)$ to denote the particle space-time position and four-momentum. The longitudinal $(z)$ direction is identified with the beam axis. The transverse momentum is $\pT = \sqrt{p_x^2 + p_y^2}$ and particles are always on the mass shell,  $E_p = \sqrt{m^2 + \pT^2 + \pL^2}$. The scalar product of two four-vectors is  $a^\mu b_\mu = a^\mu  g_{\mu\nu} b^\nu \equiv a \cdot b$ where $g^{\mu\nu} = {\rm diag} \left(1,-1,-1,-1\right)$ is the metric tensor.  For the partial derivative we use the notation $\p_\mu\equiv \p/\p x^\mu$. Throughout the paper we use natural units with $c=k_B=\hbar=1$.  The notation for various physical quantities and special functions is kept the same as in~\cite{Florkowski:2017jnz}.

%
\section{Kinetic equations}
\label{sec:ke}
%
Our starting point are three coupled relativistic Boltzmann transport equations for quark, antiquark, 
and gluon phase-space distribution functions~$ f_{\s}(x,p)$
~\cite{Florkowski:2012as,Florkowski:2013uqa,Florkowski:2014txa,Florkowski:2015cba,Florkowski:2017jnz},
\beal{kineq}
\lp p\cdot\p \rp  f_{\s}(x,p) &=&  {\cal C}\lsb  f_{\s}(x,p)\rsb,\quad \s=\Qp, \Qm,G. 
\eeal
The collisional kernel ${\cal C}$ in \rfn{kineq} is used in the relaxation time approximation~\cite{Bhatnagar:1954zz,Anderson:1974a,Anderson:1974b,Czyz:1986mr} 
\beal{colker}
{\cal C}\lsb f_\s(x,p)\rsb &=&   \lp p\cdot U \rp \frac{f_{\s, \eq}(x,p)-f_\s(x,p)}{\teq},
\eeal
where $\teq$ is the relaxation time and the four-vector $U(x)$ describes the hydrodynamic flow defined in the Landau hydrodynamic frame~\cite{LLfluid}.
For one-dimensional boost-invariant systems the structure of $U^\mu(x)$ is determined by the symmetry arguments (see Eq.~\rfn{Ubinv} below). 
%
\subsection{Equilibrium distributions}
\label{ssec:ed}
%
In \EQ{colker} the functions $f_{\s, \eq}(x,p)$ are equilibrium distribution functions, which take the Fermi-Dirac and Bose-Einstein forms for (anti)quarks and gluons, respectively,
\bea
f_{\Qpm, \eq}(x,p) &=& h^{+}_\eq\lp\f{  p\cdot U  \mp \mu }{T }\rp ,
\label{Qeq} \\
f_{\G, \eq}(x,p) &=& h^{-}_\eq\lp\f{p\cdot U}{T }\rp.
\label{Geq}
\eea
Here $T(x)$ is the effective temperature, $\mu(x)$ is the effective baryon chemical potential of quarks, and
\beal{feq}
h^{\pm}_\eq(a) &=&   \lsb\VP\exp(a) \pm 1 \rsb^{-1},
\eeal
where the sign $+1$ ($-1$) corresponds to fermions (bosons). The same value of $T(x)$ appearing in Eqs.~(\ref{Qeq}) and (\ref{Geq}), as well as the same value of $\mu(x)$ appearing in the quark and antiquark distributions, means that all particles evolve toward the same local equilibrium defined by $T(x)$ and $\mu(x)$.  Since the baryon number of quarks is 1/3, the baryon chemical potential $\mu_B$ is defined by the expression
\bel{muB}
\mu= \f{\mu_B}{3}.
\eel
%
%
\subsection{Anisotropic distributions}
\label{ssec:ad}
%
In this paper, following the main ideas of anisotropic hydrodynamics \cite{Florkowski:2010cf,Martinez:2010sc}, we make an assumption that the exact solutions, $f_{\s}(x,p)$, of \rfm{kineq} are very well approximated by the 
 Romatschke-Strickland (RS) anisotropic distributions~\cite{Romatschke:2003ms}. In the covariant version for the Bjorken expansion  they read~\cite{Florkowski:2012as}
\begin{eqnarray}
f_{\Qpm, \an}(x,p) \!\!\!&=&\!\!\! h^{+}_\eq\lp\f{\sqrt{\lp p \cdot U\rp^2+\xi_\Q  \lp p \cdot Z\rp^2} \mp \lambda }{\Lambda_\Q}\rp,
\,\,\,\,  \label{Qa} \\
f_{\G, \an}(x,p) \!\!\!&=&\!\!\! h^{-}_\eq\lp\f{\sqrt{\lp p \cdot U\rp^2+\xi_\G  \lp p \cdot Z\rp^2}}{\Lambda_G}\rp,
\label{Ga}
\end{eqnarray}
where $\xi_\Q(x)=\xi_\Qp(x)=\xi_\Qm(x)$ is the quark anisotropy parameter, $\Lambda_\Q(x)=\Lambda_\Qp(x)=\Lambda_\Qm(x)$ is the quark transverse-momentum scale, and $\lambda(x)$ is the non-equilibrium baryon chemical potential of quarks. Similarly,
$\xi_G(x)$ is the gluon anisotropy parameter and $\Lambda_G(x)$ is the gluon transverse-momentum scale.

The use of the RS ansatz for the quark and gluon distributions means that we deal with seven unknown functions: $\xi_\Q, \Lq, \xi_\G, \Lg, \lambda, \mu$, and $T$. Their spacetime dependence will be determined by using a properly selected set of moments of Eqs.~\rfn{kineq}. In this work we follow Ref.~\cite{Florkowski:2015cba} and use two equations constructed from the zeroth moments, one from the first moment, and two from the second moments. In addition, we use two so-called Landau matching conditions that guarantee the baryon number and energy-momentum conservation~\footnote{The Landau matching conditions for baryon number and four-momentum follow also from appropriate combinations of the zeroth and first moments of the kinetic equations~\rfn{kineq}, respectively.}.

%
\subsection{Boost-invariance and the tensorial basis}
\label{ssec:bitb}
%
In the transversely-homogeneous and boost-invariant case studied here the tensorial basis used in the calculations has the following form~\cite{Florkowski:2011jg}
\begin{eqnarray}
 U^\mu &=& (t/\tau,0,0,z/\tau),   \label{Ubinv} \\
 X^\mu &=& (0,1,0,0),   \label{Xbinv} \\
 Y^\mu &=& (0,0,1,0),  \label{Ybinv}  \\
 Z^\mu &=& (z/\tau,0,0,t/\tau), \label{Zbinv}
\end{eqnarray}
where $\tau$ is the (longitudinal) proper time
\begin{eqnarray}
\tau = \sqrt{t^2-z^2}.
\end{eqnarray}
The four-vectors $X^\mu$ and $Y^\mu$ will be used in calculations involving the second moment of the kinetic equations in Sec.~\ref{sect:2nd}. We note that for
one-dimensional, boost-invariant expansion the functions $\xi_\Q, \Lq, \xi_\G, \Lg, \lambda, \mu$, and $T$ may depend 
only on the proper time $\tau$.

\section{Basic observables}
\label{sec:ke}

In our calculations, all particles are assumed to be on the mass shell, \mbox{$p^2 =p \cdot p=m^2$}, so that the invariant momentum measure is
\bea
 \int dP (\ldots)  \equiv  2  \int  d^4 p \, \Theta(p^0) \delta(p^2-m^2) (\ldots)
 =    \int  \frac{d^3p}{E_p} (\ldots),
\eea
where $\Theta$ is the Heaviside step function.  Hereafter, the gluons are treated as massless, while quarks have a finite constant mass $m$.

The first, second, and third moments of the distribution functions (multiplied by $k_\s$) read 
\bea 
N_\s^\mu(x)  &=& k_\s   \int \!dP\, p^{\mu} f_\s(x,p),
\label{ncurrs} \\
T_\s^{\mu\nu}(x) &=&   k_\s \int \!dP\, p^{\mu}p^{\nu} f_\s(x,p),
\label{emtensors} \\
\Theta_\s^{\lambda\mu\nu}(x) &=&   k_\s \int \!dP\, p^{\lambda} p^{\mu}p^{\nu} f_\s(x,p),
\label{thetatensors} 
\eea
respectively. Here $k_\s\equiv g_\s/(2\pi)^3$, with $g_{Q^\pm}=3\times2\times N_f$ and $g_{G}=8\times2$ being the internal degeneracy factors for quarks  and gluons, respectively. For quarks, in addition to the colour and spin degrees of freedom, we include the flavour degeneracy $N_f=2$. While the equations \rfn{ncurrs} and \rfn{emtensors} define the \emph{particle number current} and the \emph{energy-momentum tensor} of the species ``$\s$'', respectively, \rf{thetatensors} does not have a straithforward physics interpretation. Moreover, we define the \emph{baryon number current}
\beal{bcurr}
B^\mu(x)&\equiv& \sum_\s q_\s\, N_\s^\mu(x) = \frac{k_\Q}{3}\int \!dP\, p^{\mu} \lsb\VP f_\Qp(x,p)-f_\Qm(x,p)\rsb, 
\eeal
where $q_\s=\left\{1/3,-1/3,0\right\}$ is the baryon number of quarks, antiquarks, and gluons, respectively. 

We introduce the total particle number current and total energy-momentum tensor which are given by the sums of individual components
\bea 
N^\mu(x)&=&\sum_\s N_\s^\mu(x),
\label{totncurr} \\
T^{\mu\nu}(x)&=& \sum_\s T_\s^{\mu\nu}(x).
\label{totemtensor} 
\eea
For the equilibrium distributions~\rfn{Qeq} and \rfn{Geq}, Eqs.~(\ref{ncurrs}) and \rfn{emtensors} may be tensor-decomposed in the following way
\begin{eqnarray} 
N_{\s, \eq}^\mu(x)&=&  {\cal N}^{\s, \eq} U^\mu, \label{ncurreq} \\
T_{\s, \eq}^{\mu\nu}(x)&=& {\cal E}^{\s, \eq} U^\mu U^\nu -{\cal P}^{\s, \eq} \Delta^{\mu\nu} ,
\end{eqnarray} 
while for the anisotropic functions~\rfn{Qa} and \rfn{Ga} we find~\cite{Florkowski:2008ag,Florkowski:2017jnz}
\begin{eqnarray} 
N_{\s,\an}^\mu(x)&=&  {\cal N}^{\s,\an} U^\mu, \label{ncurran} \\
T_{\s,\an}^{\mu\nu}(x)&=& {\cal E}^{\s,\an} U^\mu U^\nu -{\cal P}^{\s,\an}_T \Delta_T^{\mu\nu} +{\cal P}^{\s,\an}_L Z^\mu Z^\nu.
\label{emtensoran} 
\end{eqnarray} 
Here $\Delta^{\mu\nu}\equiv g^{\mu\nu} -U^\mu U^\nu=-X^\mu X^\nu - Y^\mu Y^\nu-Z^\mu Z^\nu$ and $\Delta_T^{\mu\nu} = -X^\mu X^\nu - Y^\mu Y^\nu$. The operator $\Delta^{\mu\nu}$ ($\Delta_T^{\mu\nu}$) projects on the space orthogonal to $U$ ($U$ and $Z$). 
The functions ${\cal N}$, ${\cal E}$, and ${\cal P}$ are the particle density, energy density, and pressure. For the anisotropic case we differentiate between
the longitudinal, ${\cal P}_L$, and transverse,  ${\cal P}_T$, pressures.

Analogous calculation for the second moment \rfn{thetatensors} gives 
\begin{eqnarray}  
\Theta_{\s, \eq}^{\mu\nu}(x)&=& \vartheta_U^{\s, \eq}\,  U^{\lambda} U^{\mu} U^{\nu}     \,-\,   \vartheta ^{\s, \eq}\, \left( U^{\lambda} \Delta^{\mu\nu} + U^{\mu} \Delta^{\lambda\nu}+    U^{\nu}\Delta ^{\lambda \mu}\right)
 \label{thetatensoreq}
\end{eqnarray} 
for the equilibrium distributions~\rfn{Qeq} and \rfn{Geq}, and 
\bea
\Theta_{\s, \an}^{\lambda\mu\nu} &=& \vartheta_U^{\s, \an}\,  U^{\lambda} U^{\mu} U^{\nu} \nonumber   \\ \,&-&\,   \vartheta_T^{\s, \an}\, \left( U^{\lambda} \Delta^{\mu\nu}_T + U^{\mu} \Delta^{\lambda\nu}_T+    U^{\nu}\Delta ^{\lambda \mu}_T\right)\nonumber\\
\,&+&\,  \vartheta_L^{\s, \an}\, \left( U^{\lambda}Z^{\mu} Z^{\nu} + U^{\mu}Z^{\lambda} Z^{\nu}+U^{\nu} Z^{\lambda} Z^{\mu} \right)
 \label{thetatensoran}
\eea 
for the anisotropic distributions~\rfn{Qa} and \rfn{Ga}. The expressions defining variables $\vartheta$ are presented in App.~\ref{s:sm}.
\section{Boost invariance}
\label{sect:e}

\subsection{Boost invariant Bialas-Czyz variables}
\label{sect:wv}
%
In the case of (0+1)-dimensional system exhibiting symmetries discussed in the previous section it is convenient to use the variables $w$ and $v$ which are defined as follows  \cite{Bialas:1984wv,Bialas:1987en}
\bea 
w & =& t \pL - z E_p =  - \, \tau \, p \cdot Z, \label{w}\\
v  &=& t E_p- z \pL  = \tau \, p \cdot U.  \label{v}
\eea 
Due to the fact that particles are on the mass shell, $w$ and $v$ are related by the formula
\beal{v1}
v\twpt &=&   \sqrt{w^2+\lp m^2+\pT^{\,2}\rp  \tau^2}.  
\eeal
Equations (\ref{w}) and (\ref{v}) can be inverted to express the energy and longitudinal momentum of a particle in terms of $w$ and $v$, namely,
$E_p= (vt+wz)/\tau^2$ and $\pL= (wt+vz)/\tau^2$. The  Lorentz invariant momentum-integration measure can be written now as
$dP = d^3p/E_p = dw\,d^2\pT/v$. For boost invariant systems, all scalar functions of space and time, such as the effective temperature $T$ and quark chemical potential $\mu$, may depend only on $\tau$. In addition, one can check that the phase-space distribution functions, which are Lorentz scalars, may depend only on the variables $w$, $\tau$ and $\bpT$. We use these properties in the next sections.
%

\subsection{Boost-invariant  distributions}

In what follows we assume that the distributions $f_\s(\tau,w,p_T)$ are given by the anisotropic RS forms $f_{\s,\an}(\tau,w,p_T)$ which follow from Eqs.~(\ref{Qa}) and (\ref{Ga}),
\begin{eqnarray}
&& f_{\Qpm, \an} \twpt = 
h^{+}_\eq \lp\f{\sqrt{\lp1+\xi_\Q\rp\lp\frac{w}{\tau}\rp^2+  m^2+\pT^2   } \mp \lambda }{\Lambda_\Q}   \rp  ,
\label{BIQan} \\
&& f_{\G, \an} \twpt = h^{-}_\eq\lp\f{\sqrt{\lp1+\xi_\G\rp\lp\frac{w}{\tau}\rp^2+\pT^2   }}{\Lambda_\G} \,  \rp . \nn \\
\label{BIGan}
\end{eqnarray}
Equations~\rfn{BIQan} and \rfn{BIGan} with $\xi_\s^0=\xi_\s(\tau_0)$, $\Lambda_\s^0=\Lambda_\s(\tau_0)$,  and $\lambda^0=\lambda(\tau_0)$ specify also our initial conditions. The boost-invariant forms of the equilibrium functions are obtained by taking the limit $\xi_\s \to 0$ in ~\rfn{BIQan} and \rfn{BIGan}.

\section{Zeroth moments of the kinetic equations}
\label{sect:motke}

The zeroth moments of the kinetic equations \rfn{kineq} give three scalar equations
\beal{0mq}
\p_\mu  \left( {\cal N}_{\Qpm, \rm a} U^\mu \right)   &=& \frac{{\cal N}_{\Qpm, \rm eq} -{\cal N}_{\Qpm, \rm a} }{\teq},
\eeal
\beal{0mg}
\p_\mu  \left( {\cal N}_{\G, \rm a} U^\mu \right)   &=& \frac{{\cal N}_{\G, \rm eq} -{\cal N}_{\G, \rm a} }{\teq}.
\eeal
To formulate the hydrodynamic framework, we cannot use all the equations listed in (\ref{0mq}) and (\ref{0mg}),  since this would lead to the overdetermined system \footnote{For a discussion of this point see Ref.~\cite{Florkowski:2015cba}.}. Therefore, we use only two equations constructed as linear combinations of (\ref{0mq}) and (\ref{0mg}). The first equation is obtained from the difference of the quark  ($\Qp$) and antiquark ($\Qm$) components in Eqs.~(\ref{0mq}),
\beal{0m1}
\f{d}{d\tau}  \left( {\cal N}_{\Qp, \rm a} - {\cal N}_{\Qm, \rm a}  \right)   + \f{{\cal N}_{\Qp, \rm a} - {\cal N}_{\Qm, \rm a}
}{\tau}
&=& \frac{{\cal N}_{\Qp, \rm eq} -{\cal N}_{\Qm, \rm eq} - \left( {\cal N}_{\Qp, \rm a} - {\cal N}_{\Qm, \rm a} \right) }{\teq},
\eeal
while the second equation is a linear combination of Eqs.~(\ref{0mq})~and~(\ref{0mg}),
\begin{eqnarray}
&& \alpha \left( \f{d {\cal N}_{Q, \rm a}}{d\tau} +  \f{{\cal N}_{\Q, \rm a}}{\tau} \right) 
+ (1-\alpha) \left( \f{d {\cal N}_{G, \rm a}}{d\tau} +  \f{ {\cal N}_{\G, \rm a}}{\tau} \right) \nn \\
&& = \alpha  \,  \frac{{\cal N}_{\Q, \rm eq}  -  {\cal N}_{\Q, \rm a}  }{\teq}
+ (1-\alpha) \,  \frac{{\cal N}_{\G, \rm eq}  -  {\cal N}_{\G, \rm a}  }{\teq}. \label{0m2}
\end{eqnarray}
The parameter $\alpha$ is a constant taken from the range $0 \leq \alpha \leq 1$. Sums of the contributions from both quarks and antiquarks are denoted by the symbol $\Q$, for example
\bel{Q}
N_{\Q, \an}^\mu = N_{\Qm, \an}^\mu + N_{\Qp, \an}^\mu, \quad {\cal N}_{\Q, \an} = {\cal N}_{\Qm, \an} + {\cal N}_{\Qp, \an}.
\eel

By performing comparisons between
the predictions of kinetic theory and the results obtained with aHydro one can check which value of $\alpha$ is optimal.
In Ref.~\cite{Florkowski:2015cba} we found that the best cases corresponded to either $\alpha=1$ or $\alpha=0$. One may
understand this behaviour, since such values of $\alpha$ do not introduce any direct coupling between the quark and gluon sectors
except for that included by the energy-momentum conservation, which is accounted for by the first moment --- such situation
takes place in the case where kinetic equations are treated exactly. Moreover, our present investigations of more complex systems also 
favour the value $\alpha=1$. We return to discussion of this point in Sec.~\ref{sect:results}.

%
\subsection{Baryon number conservation}
\label{sect:bnc}

Equation \rfn{0m1} leads directly to the constraint on the baryon number density
\bel{Ba1}
\f{d{\cal B}_{\rm a}(\tau)}{d\tau} +\f{{\cal B}_{\rm a}(\tau)}{\tau} = \f{{\cal B}_{\rm eq} - {\cal B}_{\rm a}}{\teq}.
\eel
The conservation of the baryon number requires that both the left- and the right-hand sides of \rfn{Ba1} vanish.
This leads to two equations
\bel{Ba2}
{\cal B}_{\rm a}(\tau) = \f{{\cal B}_{\rm 0} \tau_0}{\tau}
\eel
and
\bel{Ba3}
{\cal B}_{\rm a}(\tau)  = {\cal B}_{\rm eq}(\tau),
\eel
which gives explicitly
\begin{eqnarray}
\frac{ \Lambda_\Q^3}{ \sqrt{1+\xi_\Q}}  
\sinh\lp\frac{\lambda}{\Lambda_\Q}\rp\,{\cal H}_{\cal B}\lp \frac{m}{\Lambda_\Q},   \frac{\lambda}{\Lambda_\Q}\rp
= T^3 \sinh\lp\frac{\mu}{T}\rp\,{\cal H}_{\cal B}\lp \frac{m}{T},   \frac{\mu}{T}\rp
\label{BaBeq}
\end{eqnarray}
and
\begin{eqnarray}
\frac{16 \pi k_\Q T^3}{3} \sinh\lp\frac{\mu}{T}\rp\,{\cal H}_{\cal B}\lp \frac{m}{T},   \frac{\mu}{T}\rp = \f{{\cal B}_{\rm 0} \tau_0}{\tau}.
\label{BaB0}
\end{eqnarray}
The function ${\cal H}_{\cal B}$ is defined explicitly in Appendix A1 of Ref.~\cite{Florkowski:2017jnz}. 

The last two equations can be  used to determine $\lambda$ and $\mu$ in terms of ${\cal B}_{\rm 0} \tau_0/\tau$, $T$, $\Lambda_\Q$,
and $\xi_\Q$. Thus, in the following equations we may treat $\lambda$ and $\mu$ as known functions of other hydrodynamic
variables.~\footnote{In the case of classical statistics, the function ${\cal H}_{\cal B}$ becomes independent of the second argument 
and Eqs.~\rfn{BaBeq} and \rfn{BaB0} can be easily solved for $\mu$ and $\lambda$. However,  in the general case of Fermi-Dirac statistics one has to solve
 Eqs.~\rfn{BaBeq} and \rfn{BaB0} numerically, together with other hydrodynamic equations. }

%
\subsection{Sum of the zeroth moments}
\label{sect:bnc}

Equation \rfn{0m2} can be written in the form
\begin{eqnarray}
&& \f{d}{d\tau} \left[ \alpha \Lambda_\Q^3 \left(  \tilde{{\cal H}}_{\cal N}^+\lp \f{1}{\sqrt{1+\xi_\Q}}, \frac{m}{\Lambda_\Q}, - \frac{\lambda}{\Lambda_\Q}\rp
+   \tilde{{\cal H}}_{\cal N}^+\lp \f{1}{\sqrt{1+\xi_\Q}}, \frac{m}{\Lambda_\Q}, + \frac{\lambda}{\Lambda_\Q}\rp  \right) \right. \nn \\
&& \hspace{3cm} \left. + \, (1-\alpha) \, r \,  \Lambda_\G^3 \tilde{{\cal H}}_{\cal N}^-\lp \f{1}{\sqrt{1+\xi_\G}}, 0,0\rp   \right] \nn \\
&& + \left(\f{1}{\tau} + \f{1}{\teq} \right) \left[ \alpha \Lambda_\Q^3 \left(  \tilde{{\cal H}}_{\cal N}^+\lp \f{1}{\sqrt{1+\xi_\Q}}, \frac{m}{\Lambda_\Q}, - \frac{\lambda}{\Lambda_\Q}\rp
+   \tilde{{\cal H}}_{\cal N}^+\lp \f{1}{\sqrt{1+\xi_\Q}}, \frac{m}{\Lambda_\Q}, + \frac{\lambda}{\Lambda_\Q}\rp  \right) \right. \nn \\
&& \hspace{3cm} \left. + \, (1-\alpha) \, r \,  \Lambda_\G^3 \tilde{{\cal H}}_{\cal N}^-\lp \f{1}{\sqrt{1+\xi_\G}}, 0,0\rp   \right] \nn \\
&=& \f{T^3}{\teq} \left[ \alpha \left( \tilde{{\cal H}}_{\cal N}^+\lp 1, \frac{m}{T}, - \frac{\mu}{T}\rp
+   \tilde{{\cal H}}_{\cal N}^+\lp 1, \frac{m}{T}, + \frac{\mu}{T}\rp \right) + \, (1-\alpha) \, r \,  \tilde{{\cal H}}_{\cal N}^-\lp 1, 0,0\rp   \right] .
\label{ZM3}
\end{eqnarray}
Here we have introduced the ratio of the internal degeneracies
\begin{equation}
r = \f{k_\G }{k_\Q} = \f{g_\G }{g_\Q} = \f{4}{3}.
\end{equation}
The functions $ \tilde{{\cal H}}_{\cal N}^\pm$ are defined explicitly in Appendix A1 of Ref.~\cite{Florkowski:2017jnz}.

\section{First moments of the kinetic equations}
\label{sect:motke}
By considering sum over ``s'' of the first moments of the kinetic equations \rfn{kineq} we have 

\beal{KEfirsthmom3}
\p_\mu  {{T}}^{\mu\nu}_{\an}  &=& U_\mu   \frac{{T}^{\mu\nu}_\eq -{T}^{\mu\nu}_{\an }}{\teq}.
\eeal
The energy-momentum conservation requires that the right-hand side of Eq.~(\ref{KEfirsthmom3}) vanishes, which leads to the Landau matching for the energy density 
\begin{equation}
{\cal E}^{\an }  = {\cal E}^\eq,
\label{energyLMC}
\end{equation} 
where ${\cal E}^{\an } = {\cal E}^{\Q,\an}  + {\cal E}^{\G, \an}$  and ${\cal E}^\eq = {\cal E}^{\Q, \eq} + {\cal E}^{G, \eq}$ contain contributions from quarks, antiquarks, and gluons. In the explicit notation we obtain
\begin{eqnarray}
 &&  \Lambda_\Q^4  \left( \tilde{{\cal H}}^+\lp \f{1}{\sqrt{1+\xi_\Q}}, \frac{m}{\Lambda_\Q}, - \frac{\lambda}{\Lambda_\Q}\rp
 +  \tilde{{\cal H}}^+\lp \f{1}{\sqrt{1+\xi_\Q}}, \frac{m}{\Lambda_\Q}, + \frac{\lambda}{\Lambda_\Q}\rp \right) \nn \\
 && \hspace{2cm} + \, r \, \Lambda_\G^4 \tilde{{\cal H}}^-\lp \f{1}{\sqrt{1+\xi_\G}}, 0,0\rp \nn \\
 &&   =  T^4 \left( \tilde{{\cal H}}^+\lp 1, \frac{m}{T}, - \frac{\mu}{T}\rp +  \tilde{{\cal H}}^+\lp 1, \frac{m}{T}, + \frac{\mu}{T}\rp
 + r \,  \tilde{{\cal H}}^-\lp 1, 0,0\rp   \right), 
 \label{FM1}
\end{eqnarray}
where the functions $ \tilde{{\cal H}}^\pm$ are defined explicitly in Appendix A1 of Ref.~\cite{Florkowski:2017jnz}.

On the other hand, the left hand side of Eq.~(\ref{KEfirsthmom3}) leads to the equation expressing the energy-momentum conservation of the form
\bel{SECOND-EQUATION-1}
\f{d{\cal E}^{\an }   }{d\tau} = -\f{{\cal E}^{\an }   + {\cal P}^{\an } _{L}  }{\tau},
\eel
where ${\cal P}^{\an }_L  = {\cal P}_L^{\Q,\an} + {\cal P}_L^{G,\an}$ is the total longitudinal momentum of the system. This leads to the equation
\begin{eqnarray}
&& \f{d}{d\tau} \left[ \Lambda_\Q^4  \left( \tilde{{\cal H}}^+\lp \f{1}{\sqrt{1+\xi_\Q}}, \frac{m}{\Lambda_\Q}, - \frac{\lambda}{\Lambda_\Q}\rp
 +  \tilde{{\cal H}}^+\lp \f{1}{\sqrt{1+\xi_\Q}}, \frac{m}{\Lambda_\Q}, + \frac{\lambda}{\Lambda_\Q}\rp \right) \right. \nn \\
&& \left. \hspace{2cm}  + \, r \, \Lambda_\G^4 \tilde{{\cal H}}^-\lp \f{1}{\sqrt{1+\xi_\G}}, 0,0\rp \right] \nn \\
&& = -\f{1}{\tau}  \left[ \Lambda_\Q^4  \left( \tilde{{\cal H}}^+\lp \f{1}{\sqrt{1+\xi_\Q}}, \frac{m}{\Lambda_\Q}, - \frac{\lambda}{\Lambda_\Q}\rp
 +  \tilde{{\cal H}}^+\lp \f{1}{\sqrt{1+\xi_\Q}}, \frac{m}{\Lambda_\Q}, + \frac{\lambda}{\Lambda_\Q}\rp \right) \right. \nn \\
 &&  \hspace{1.2cm} +  \Lambda_\Q^4  \left( \tilde{{\cal H}}^+_L \lp \f{1}{\sqrt{1+\xi_\Q}}, \frac{m}{\Lambda_\Q}, - \frac{\lambda}{\Lambda_\Q}\rp
 +  \tilde{{\cal H}}^+_L \lp \f{1}{\sqrt{1+\xi_\Q}}, \frac{m}{\Lambda_\Q}, + \frac{\lambda}{\Lambda_\Q}\rp \right)  \nn \\
&& \left. \hspace{2cm}  + \, r \, \Lambda_\G^4 \left( \tilde{{\cal H}}^-\lp \f{1}{\sqrt{1+\xi_\G}}, 0,0\rp
+ \tilde{{\cal H}}^-_L \lp \f{1}{\sqrt{1+\xi_\G}}, 0,0\rp \right) \right], \label{FM2} 
\end{eqnarray}
where the functions $ \tilde{{\cal H}}_L^\pm$ are again defined in Appendix A1 of Ref.~\cite{Florkowski:2017jnz}. We note that Eqs.~\rfn{FM1} and \rfn{FM2} couple the quark and gluon hydrodynamic parameters in the way similar to that known from the exact treatment of the kinetic equations.

\section{Second moments of the kinetic equations}
\label{sect:2nd}

In order to close the system of dynamical equations we finally consider the second moments of  the kinetic equations \rfn{kineq}, 

\beal{KEfirsthmom4}
\p_\lambda  {{\Theta}}_{\s, \an}^{\lambda\mu\nu}  &=& U_\lambda   \frac{{\Theta}^{\lambda\mu\nu}_{\s,\eq} -{\Theta}^{\lambda\mu\nu}_{\s, \an}}{\teq}.
\eeal
Using tensor decompositions \rfn{thetatensoreq} and \rfn{thetatensoran} and performing the projections of Eqs.~(\ref{KEfirsthmom4}) on the basis four-vectors, one obtains however an overdetermined system of equations. A possible remedy to this problem was proposed in Ref.~\cite{Tinti:2013vba} where a selection rule for equations was proposed, see also \cite{Nopoush:2014pfa}.
In this work we select the same combinations of the second moments of the Boltzmann equations as in Ref.~\cite{Florkowski:2015cba}, which follows methodology of Refs.~\cite{Tinti:2013vba, Nopoush:2014pfa}. This implies that we use the equations of the form
\begin{eqnarray}
&& \frac{d}{d\tau}\ln\vartheta_T^{\Q,\an}-\frac{d}{d\tau}\ln\vartheta_L^{\Q,\an}-\frac{2}{\tau}
=\frac{\vartheta^{\Q,\eq}}{\tau_{\rm eq}}\left[\frac{1}{\vartheta_T^{\Q,\an}}-\frac{1}{\vartheta_L^{\Q,\an}}  \right] , \label{sumXq} \\
&& \frac{d}{d\tau}\ln\vartheta_T^{\G,\an}-\frac{d}{d\tau}\ln\vartheta_L^{\G,\an}-\frac{2}{\tau}
=\frac{\vartheta^{\G,\eq}}{\tau_{\rm eq}}\left[\frac{1}{\vartheta_T^{\G,\an}}-\frac{1}{\vartheta_L^{\G,\an}}  \right] .  \label{sumXg}
\end{eqnarray}
It has been demonstrated in Ref.~\cite{Tinti:2013vba} that such forms are consistent with the Isreal-Stewart theory for systems being close to local equilibrium.
 The explicit expressions for the functions $\vartheta$ are given in the Appendix \ref{s:sm}.
Using~Eqs.~\rfn{sumXq} and \rfn{sumXg} we find

\begin{eqnarray}
&& \f{1}{1+\xq} \f{d\xq}{d\tau} -\frac{2}{\tau}
= - \frac{\xq  (1+\xq)^{1/2}}{\tau_{\rm eq}} \frac{T^5 }{\Lq^5} \f{\tilde{{\cal H}}_{\vartheta}^+\lp  \frac{m}{T}, - \frac{\mu}{T}\rp +\tilde{{\cal H}}_{\vartheta}^+\lp  \frac{m}{T}, + \frac{\mu}{T}\rp} {\tilde{{\cal H}}_{\vartheta}^+\lp  \frac{m}{\Lq}, - \frac{\lambda}{\Lq}\rp +\tilde{{\cal H}}_{\vartheta}^+\lp  \frac{m}{\Lq}, + \frac{\lambda}{\Lq}\rp}
, \label{sumXqgen} \\
&&  \f{1}{1+\xg} \f{d\xg}{d\tau} -\frac{2}{\tau}
= - \frac{\xg  (1+\xg)^{1/2}}{\tau_{\rm eq}} \f{T^5} {\Lg^5}, \label{sumXggen}
\end{eqnarray} 
with the function $\tilde{{\cal H}}_{\vartheta}^+$ defined by \rf{Htheta}.

\begin{figure}[t!]
\includegraphics[angle=0,width=0.6\textwidth]{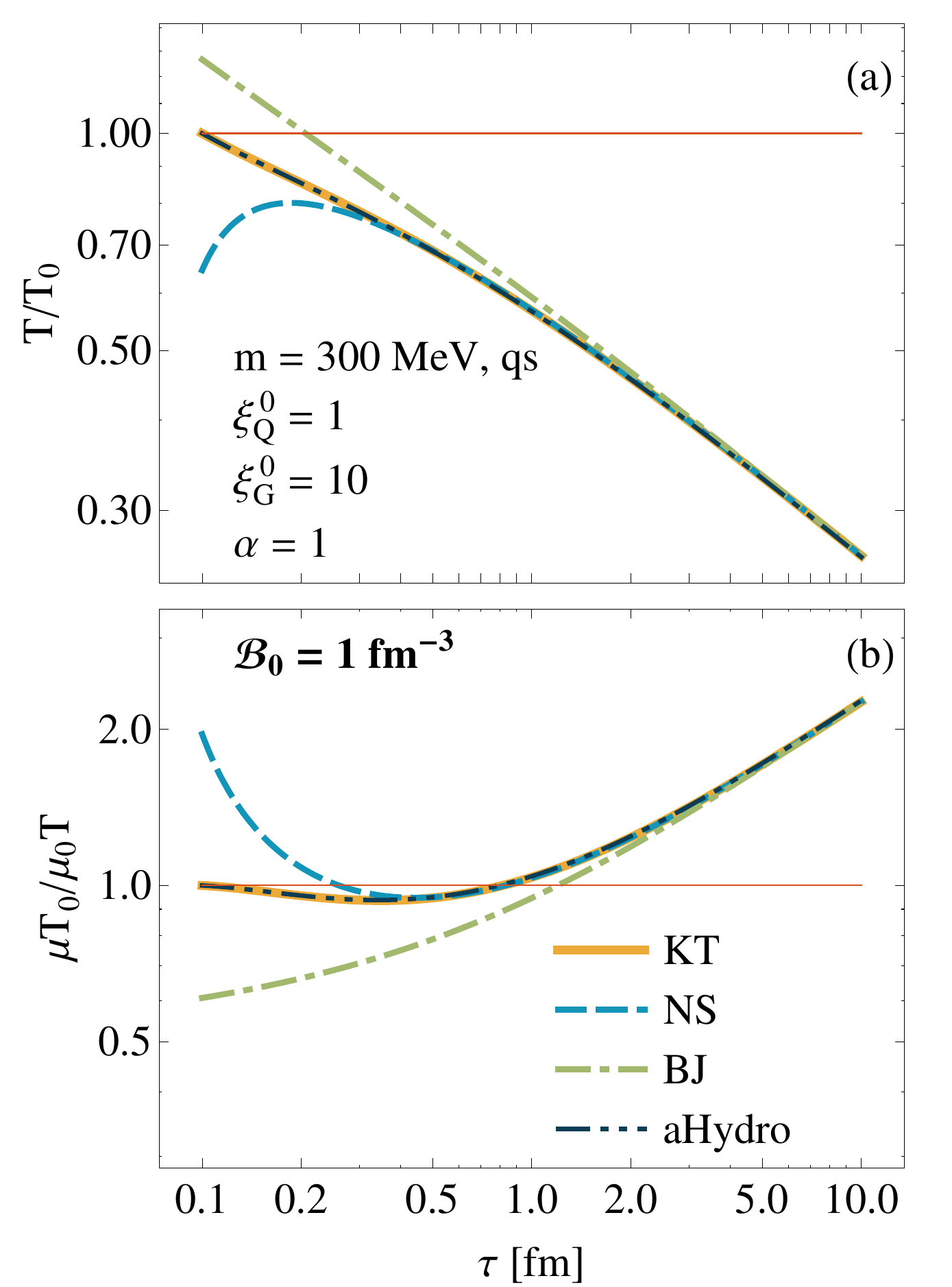} 
\caption{(Color online)  Proper-time dependence of the effective temperature $T$, panel (a), and the effective baryon chemical potential $\mu$ divided by $T$, panel (b), both rescaled by the initial values.  The exact kinetic-theory result (KT, orange solid line) is compared with the aHydro (double-dot-dashed line), Navier-Stokes (NS, blue dashed line), and perfect-fluid (BJ, green dot-dashed line) results, respectively. The label ``qs'' denotes using the quantum statistics for both quarks and gluons. 
}
\label{fig:MuT_oo_ex_Bj_NS_aH}
\end{figure}
%

\section{results}
\label{sect:results}

Equations \rfn{BaBeq}, \rfn{BaB0}, \rfn{ZM3}, \rfn{FM1}, \rfn{FM2}, \rfn{sumXqgen} and \rfn{sumXggen} are seven equations for
seven unknown functions of the proper time: $\xi_\Q(\tau), \Lq(\tau), \xi_\G(\tau), \Lg(\tau), \lambda(\tau), \mu(\tau)$, and $T(\tau)$.
Three equations [\rfn{BaBeq}, \rfn{BaB0} and \rfn{FM1}] are algebraic but one can differentiate them with respect to proper time and
use them together with the remaining four equations [\rfn{ZM3},  \rfn{FM2}, \rfn{sumXqgen} and \rfn{sumXggen}] as a system of 
seven ordinary differential equations of the first order. Such equations require initial values which we choose in the similar way as
in~\cite{Florkowski:2017jnz}. The initial values of the anisotropy parameters correspond to the two options:  i) $\xi_{\Q}^{0}=1$ and $\xi_{\G}^{0}=10$, and 
ii)~$\xi_{\Q}^{0}=-0.5$ and $\xi_{\G}^{0}=-0.25$. Such values define oblate-oblate and prolate-prolate initial momentum distributions of quarks and gluons, respectively. The same initial values for $\xi_{\Q}^{0}$ and $\xi_{\G}^{0}$ were used before in Refs.~\cite{Florkowski:2015cba,Florkowski:2017jnz}. The initial transverse momentum scales of quarks and gluons are assumed to be the same and equal to $\Lambda_{\Q}^{0}=\Lambda_{\G}^{0}=1$~GeV.  The gluons are treated as massless, while quarks have a finite mass of 300~MeV. The initial non-equilibrium  chemical potential $\lambda_0$ is chosen in such a way that the initial baryon number density is ${\cal B}_0=$~0.001~fm$^{-3}$ or ${\cal B}_0=$~1~fm$^{-3}$.  The initial proper time is $\tau_0=$~0.1~fm and the relaxation time is $\teq=$~0.25~fm. The results shown in this section were obtained with $\alpha=1$ used in \rf{0m2}. We comment on this choice below. 

\begin{figure}[t!]
\includegraphics[angle=0,width=0.9\textwidth]{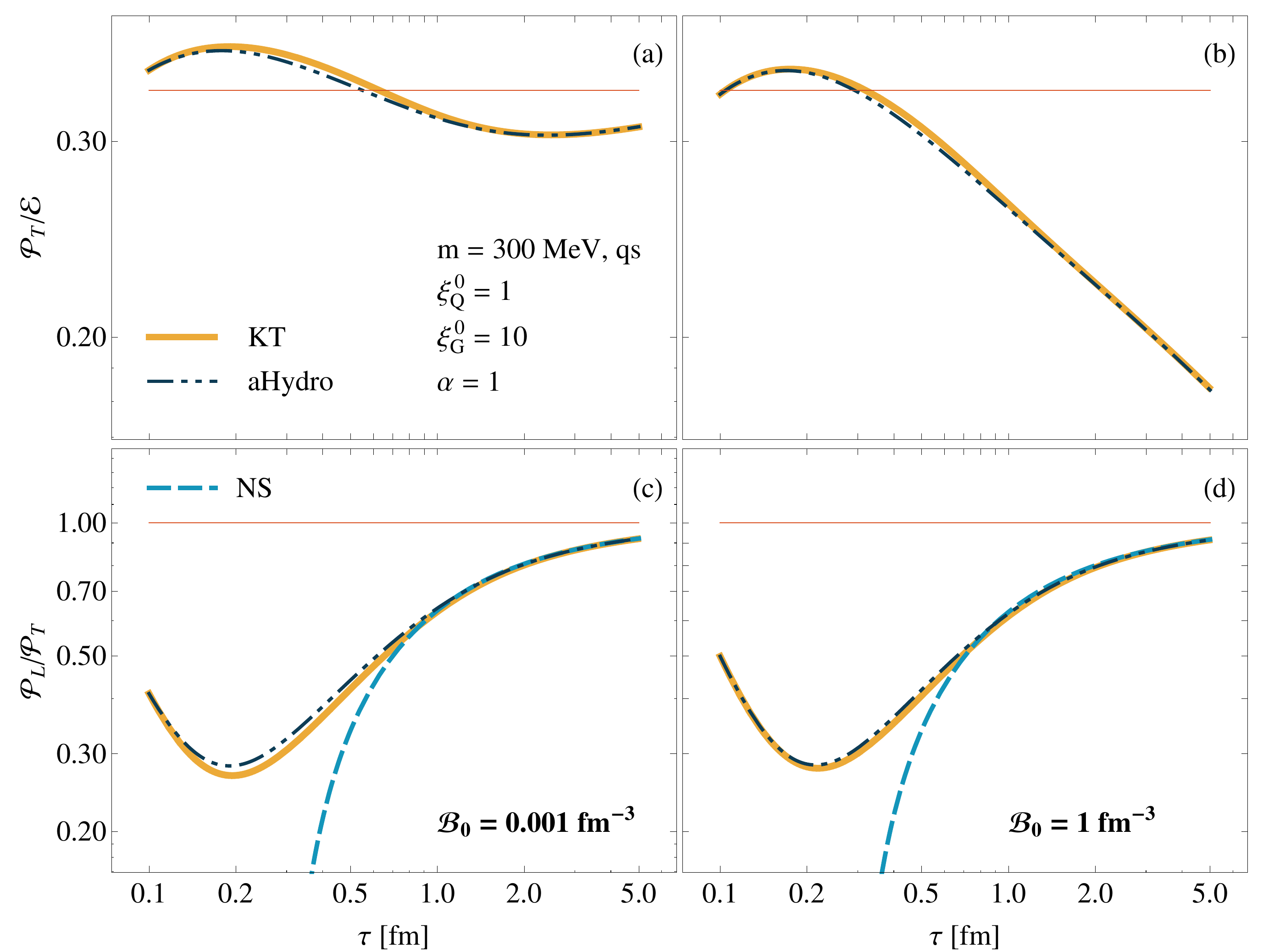} 
\caption{(Color online) Proper-time dependence of the transverse pressure over energy density ratio
[${\cal P}_T/{\cal E}$,  panels (a) and (b)] and the longitudinal pressure over transverse pressure ratio [${\cal P}_L/{\cal P}_T$, panels  (c) and (d)] for the initial oblate-oblate configuration. Notation the same as in Fig.~\ref{fig:MuT_oo_ex_Bj_NS_aH}.}
\label{fig:P_m300qs_NS_KT_aH}
\end{figure}

In Fig.~\ref{fig:MuT_oo_ex_Bj_NS_aH} we show the proper-time dependence of the effective temperature $T$, panel (a), and the effective baryon chemical potential $\mu$ divided by $T$, panel (b).  The exact kinetic-theory result (KT, orange solid line) is compared with the aHydro (double-dot-dashed line), Navier-Stokes (NS, blue dashed line), and perfect-fluid (BJ, green dot-dashed line) results, respectively. The kinetic and aHydro calculations start with the same initial conditions corresponding to oblate-oblate configuration defined above. The NS and BJ calculations are adjusted in such a way as to reproduce the late time behaviour of the KT calculation.
We observe that the KT and aHydro results agree very well during the whole evolution process. On the other hand, the NS and BJ results can reproduce
the KT result only when the system approaches local equilibrium. As expected, the NS framework coincides with the KT result much earlier than~BJ, since NS includes non-equilibrium, viscous corrections.

\begin{figure}[t!]
\includegraphics[angle=0,width=0.9\textwidth]{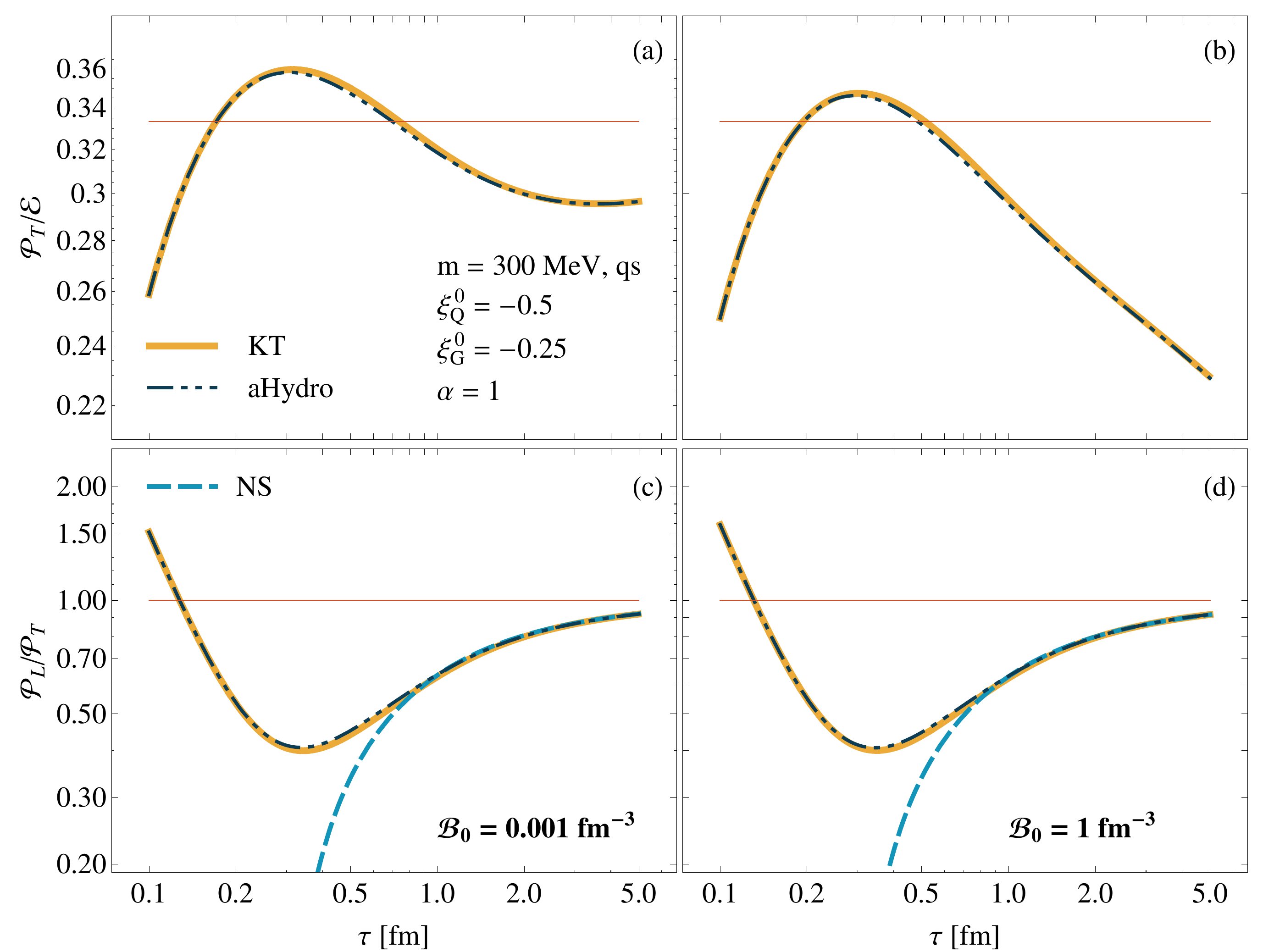} 
\caption{(Color online) Same as Fig.~\ref{fig:P_m300qs_NS_KT_aH} but for the prolate-prolate initial configuration. }
\label{fig:P_m300qs_pp_NS_KT_aH}
\end{figure}
%

If the functions $\xi_\Q(\tau), \Lq(\tau), \xi_\G(\tau), \Lg(\tau), \lambda(\tau), \mu(\tau)$, and $T(\tau)$ are known, we may determine  the anisotropic RS distribution functions for quarks and gluons and, subsequently, use these distributions to calculate various physical observables. Explicit expressions for the energy density and
different pressure components are given in~\cite{Florkowski:2017jnz}. In Fig.~\ref{fig:P_m300qs_NS_KT_aH} we show the proper-time dependence of the transverse pressure over energy density ratio [${\cal P}_T/{\cal E}$,  panels (a) and (b)] and the longitudinal pressure over transverse pressure ratio [${\cal P}_L/{\cal P}_T$,  panels (c) and (d)] for the initial oblate-oblate configuration. The left (right) panels describe the case ${\cal B}_0=$~0.001~fm$^{-3}$  (${\cal B}_0=$~1~fm$^{-3}$). We find that aHydro reproduces very well the kinetic-theory results. Similar, very good agreement between the KT and aHydro results is shown in Fig.~\ref{fig:P_m300qs_pp_NS_KT_aH} for the prolate-prolate initial conditions with $\xi_{\Q}^{0}=-0.5$ and $\xi_{\G}^{0}=-0.25$. In the lower panels of Figs.~\ref{fig:P_m300qs_NS_KT_aH} and \ref{fig:P_m300qs_pp_NS_KT_aH} we demonstrate that the ${\cal P}_T/{\cal P}_T$ ratio approaches always the NS result.

Other studied cases, not shown here, also confirm very good performance of anisotropic hydrodynamics as an approximation for the kinetic theory (as far as the observables studied in this work are considered). We stress that we have achieved this agreement with the parameter $\alpha$ set equal to unity in  \rf{0m2}. Slightly worse but satisfactory agreement can be  found also in the case $\alpha =0$. On the other hand, choosing a finite value of $\alpha$ from the range $0 < \alpha < 1$ spoils the agreement. This behaviour was already observed in  \cite{Florkowski:2015cba}  where the case with massless particles and classical statistics was studied. Apparently, the values $0 < \alpha < 1$ introduce a redundant coupling between quarks and gluons, which is absent in the RTA kinetic theory.

\section{Summary}
\label{sect:sumcon}

In this work we have compared the anisotropic-hydrodynamics results for a quark-gluon fluid mixture with predictions of the RTA kinetic theory. We have generalised previous results by including the finite quark mass, the quantum statistics for both quarks and gluons, and the finite baryon number density. We have found very good agreement between the aHydro and KT results. Our study corroborates earlier assumptions made in  \cite{Florkowski:2015cba} for construction of aHydro for mixtures, namely, the use of the second moments of kinetic equations and the appropriate choice of the zeroth moments (corresponding to $\alpha=1$).
  
\begin{acknowledgments}
W.F.  and R.R. were supported in part by the Polish National Science Center Grant No. 2016/23/B/ST2/00717. 
\end{acknowledgments}

\appendix
%
\section{Third moments of the distribution function}
\label{s:sm}
%
The third moment of the distribution function is expressed by the momentum integral
\bea  
\Theta_\s^{\lambda\mu\nu}(x) &=&   k_\s \int \!dP\, p^{\lambda} p^{\mu}p^{\nu} f_\s(x,p).
\label{app:thetatensors} 
\eea
The latter may be decomposed in the tensorial basis constructed from the tensor products of the basis four-vectors 
$A^\mu_{(\alpha)}=\{U^\mu, X^\mu, Y^\mu, Z^\mu\}$,
\beal{app:thetadecomp}
\Theta_\s^{\lambda\mu\nu}(x) &=& \sum_{A,B,C} c_{ABC}^{\s}  A^\lambda B^\mu C^\nu,
\eeal
where the coefficients $c_{ABC}^{\s}$ are defined through the expression 
\beal{app:thetadecomp2}
 c_{ABC}^{\s} &=& A_{\lambda}^{\,} B_{\mu}^{\,} C_{\nu}^{\,} \Theta_\s^{\lambda\mu\nu}(x) \, A^2  B^2 C^2.
\eeal
Using \EQ{app:thetatensors} one thus has
\bea  
c_{ABC}^{\s} &=& k_\s\int \!dP\,  \lp p \cdot A \rp \lp p \cdot B \rp \lp p \cdot C \rp A^2 B^2 C^2 f_\s(x,p).
\label{app:thetadecompcoefgen}
\eeal
For the distribution functions specified in \EQSTWO{Qeq}{Geq} and \EQSTWO{Qa}{Ga}, due to the symmetry arguments, the  only non-vanishing coefficients $c_{ABC}^{\s}$ out of those in \EQ{app:thetadecompcoefgen} are those with  an even number of each spatial index of $p^{\mu}$ which means that  \EQ{app:thetadecomp} may be expressed as follows
\beal{app:thetadecomp3}
\Theta_\s^{\lambda\mu\nu}(x) &=& c_{UUU}^{\s}  U^\lambda U^\mu U^\nu   +\sum_{A} c_{UAA}^{\s}  (U^\lambda A^\mu A^\nu+ A^\lambda U^\mu A^\nu+ A^\lambda A^\mu U^\nu).
\eeal
%
\subsection{Anisotropic RS distribution}
\label{ssec:appad}
%
For the anisotropic distribution functions one may exploit the SO(2) symmetry of Eqs.~(\ref{Qa})--(\ref{Ga}) in transverse momentum plane and write
\bea
\Theta_{\s, \an}^{\lambda\mu\nu} &=& \vartheta_U^{\s, \an}\,  U^{\lambda} U^{\mu} U^{\nu}    \\ \,&-&\,  \vartheta_T^{\s, \an}\, \left( U^{\lambda} \Delta^{\mu\nu}_T + U^{\mu} \Delta^{\lambda\nu}_T+    U^{\nu}\Delta ^{\lambda \mu}_T\right)\\
\,&+&\,  \vartheta_L^{\s, \an}\, \left( U^{\lambda}Z^{\mu} Z^{\nu} + U^{\mu}Z^{\lambda} Z^{\nu}+U^{\nu} Z^{\lambda} Z^{\mu} \right),
 \label{app:thetatensoran}
\eea  
where $\vartheta_U^{\s, \an} \equiv c_{UUU}^{\s, \an}$, $
\vartheta_T^{\s, \an} \equiv c_{UXX}^{\s, \an}=c_{UYY}^{\s, \an}$, and $ 
\vartheta_L^{\s, \an} \equiv c_{UZZ}^{\s, \an}$.

For quarks explicit calculation gives 
\begin{eqnarray}
\vartheta_U^{\Qpm, \an}    &=&   \frac{4\pi k_\Q \Lambda_\Q^5}{3}   \frac{(3+2\xi_\Q)}{(1+\xi_\Q)^{3/2}} \tilde{{\cal H}}_{\vartheta}^+\lp  \frac{m}{\Lambda_\Q}, \mp \frac{\lambda}{\Lambda_\Q}\rp +m^2 {\cal N}^{\Qpm, \an} ,
\label{app:thetaU}\\
\vartheta_T^{\Qpm, \an}   &=&   \frac{4\pi k_\Q \Lambda_\Q^5}{3}   \frac{1}{\sqrt{1+\xi_\Q}} \tilde{{\cal H}}_{\vartheta}^+\lp  \frac{m}{\Lambda_\Q}, \mp \frac{\lambda}{\Lambda_\Q}\rp,
\label{app:thetaT}\\
\vartheta_L^{\Qpm,\an}    &=&   \frac{4\pi k_\Q \Lambda_\Q^5}{3}   \frac{1}{(1+\xi_\Q)^{3/2}} \tilde{{\cal H}}_{\vartheta}^+\lp  \frac{m}{\Lambda_\Q}, \mp \frac{\lambda}{\Lambda_\Q}\rp,
\label{app:thetaL}
\end{eqnarray}
where 
\beal{Htheta}
\tilde{{\cal H}}_{\vartheta}^\pm\lp y,z\rp &\equiv& \int\limits_0^\infty r^4  dr \, h^{^\pm}_\eq\lp \sqrt{r^2+y^2}+z\rp.
\eeal
Analogous expressions hold for gluons where the integral in \EQ{Htheta} yields $\tilde{{\cal H}}_{\vartheta}^\pm\lp 0,0\rp  = 24 \zeta(5)$ with $\zeta$ being the Riemann zeta function.
\subsection{Equilibrium distribution}
\label{ssec:ed}
%
For the equilibrium distribution functions  the SO(3) symmetry of the Eqs.~(\ref{Qeq})--(\ref{Geq}) in the momentum space allows one to write
\bea
\Theta_{\s, \eq}^{\lambda\mu\nu} &=& \vartheta_U^{\s, \eq}\,  U^{\lambda} U^{\mu} U^{\nu}   - \vartheta^{\s, \eq}\, \left( U^{\lambda} \Delta^{\mu\nu} + U^{\mu} \Delta^{\lambda\nu}+    U^{\nu}\Delta ^{\lambda \mu}\right),
 \label{app:thetatensoreq}
\eea  
where $\vartheta_U^{\s, \eq} \equiv c_{UUU}^{\s, \eq}$, and $
\vartheta^{\s, \eq} \equiv c_{UXX}^{\s, \eq}=c_{UYY}^{\s, \eq} = c_{UZZ}^{\s, \eq}$ are obtained from expressions (\ref{app:thetaU})--(\ref{app:thetaL}) taking the limit $\xi_s\to 0$, where $\Lambda_s\to T$.

\bibliography{hydro_review}{}

\providecommand{\href}[2]{#2}\begingroup\raggedright\begin{thebibliography}{10}

\bibitem{Florkowski:2012as}
W.~Florkowski, R.~Maj, R.~Ryblewski, and M.~Strickland, ``{Hydrodynamics of
  anisotropic quark and gluon fluids},''
  \href{http://dx.doi.org/10.1103/PhysRevC.87.034914}{{\em Phys. Rev.} {\bf
  C87} (2013) no.~3, 034914},
\href{http://arxiv.org/abs/1209.3671}{{\tt arXiv:1209.3671 [nucl-th]}}.

\bibitem{Florkowski:2013uqa}
W.~Florkowski and R.~Maj, ``{Mixture of anisotropic fluids},''
  \href{http://dx.doi.org/10.5506/APhysPolB.44.2003}{{\em Acta Phys. Polon.}
  {\bf B44} (2013) no.~10, 2003--2017},
\href{http://arxiv.org/abs/1309.2786}{{\tt arXiv:1309.2786 [nucl-th]}}.

\bibitem{Florkowski:2014txa}
W.~Florkowski and O.~Madetko, ``{Kinetic description of mixtures of anisotropic
  fluids},'' \href{http://dx.doi.org/10.5506/APhysPolB.45.1103}{{\em Acta Phys.
  Polon.} {\bf B45} (2014) no.~5, 1103--1118},
\href{http://arxiv.org/abs/1402.2401}{{\tt arXiv:1402.2401 [nucl-th]}}.

\bibitem{Florkowski:2015cba}
W.~Florkowski, E.~Maksymiuk, R.~Ryblewski, and L.~Tinti, ``{Anisotropic
  hydrodynamics for a mixture of quark and gluon fluids},''
  \href{http://dx.doi.org/10.1103/PhysRevC.92.054912}{{\em Phys. Rev.} {\bf
  C92} (2015) no.~5, 054912},
\href{http://arxiv.org/abs/1508.04534}{{\tt arXiv:1508.04534 [nucl-th]}}.

\bibitem{Tinti:2013vba}
L.~Tinti and W.~Florkowski, ``{Projection method and new formulation of
  leading-order anisotropic hydrodynamics},''
  \href{http://dx.doi.org/10.1103/PhysRevC.89.034907}{{\em Phys. Rev.} {\bf
  C89} (2014) no.~3, 034907},
\href{http://arxiv.org/abs/1312.6614}{{\tt arXiv:1312.6614 [nucl-th]}}.

\bibitem{Florkowski:2013lza}
W.~Florkowski, R.~Ryblewski, and M.~Strickland, ``{Anisotropic Hydrodynamics
  for Rapidly Expanding Systems},''
  \href{http://dx.doi.org/10.1016/j.nuclphysa.2013.08.004}{{\em Nucl. Phys.}
  {\bf A916} (2013)  249--259},
\href{http://arxiv.org/abs/1304.0665}{{\tt arXiv:1304.0665 [nucl-th]}}.

\bibitem{Florkowski:2013lya}
W.~Florkowski, R.~Ryblewski, and M.~Strickland, ``{Testing viscous and
  anisotropic hydrodynamics in an exactly solvable case},''
  \href{http://dx.doi.org/10.1103/PhysRevC.88.024903}{{\em Phys. Rev.} {\bf
  C88} (2013)  024903},
\href{http://arxiv.org/abs/1305.7234}{{\tt arXiv:1305.7234 [nucl-th]}}.

\bibitem{Bazow:2013ifa}
D.~Bazow, U.~W. Heinz, and M.~Strickland, ``{Second-order (2+1)-dimensional
  anisotropic hydrodynamics},''
  \href{http://dx.doi.org/10.1103/PhysRevC.90.054910}{{\em Phys. Rev.} {\bf
  C90} (2014) no.~5, 054910},
\href{http://arxiv.org/abs/1311.6720}{{\tt arXiv:1311.6720 [nucl-th]}}.

\bibitem{Florkowski:2014sfa}
W.~Florkowski, E.~Maksymiuk, R.~Ryblewski, and M.~Strickland, ``{Exact solution
  of the (0+1)-dimensional Boltzmann equation for a massive gas},''
  \href{http://dx.doi.org/10.1103/PhysRevC.89.054908}{{\em Phys. Rev.} {\bf
  C89} (2014) no.~5, 054908},
\href{http://arxiv.org/abs/1402.7348}{{\tt arXiv:1402.7348 [hep-ph]}}.

\bibitem{Florkowski:2014sda}
W.~Florkowski and E.~Maksymiuk, ``{Exact solution of the (0+1)-dimensional
  Boltzmann equation for massive Bose-Einstein and Fermi-Dirac gases},''
  \href{http://dx.doi.org/10.1088/0954-3899/42/4/045106}{{\em J. Phys.} {\bf
  G42} (2015) no.~4, 045106},
\href{http://arxiv.org/abs/1411.3666}{{\tt arXiv:1411.3666 [hep-ph]}}.

\bibitem{Denicol:2014tha}
G.~S. Denicol, U.~W. Heinz, M.~Martinez, J.~Noronha, and M.~Strickland,
  ``{Studying the validity of relativistic hydrodynamics with a new exact
  solution of the Boltzmann equation},''
  \href{http://dx.doi.org/10.1103/PhysRevD.90.125026}{{\em Phys. Rev.} {\bf
  D90} (2014) no.~12, 125026},
\href{http://arxiv.org/abs/1408.7048}{{\tt arXiv:1408.7048 [hep-ph]}}.

\bibitem{Denicol:2014xca}
G.~S. Denicol, U.~W. Heinz, M.~Martinez, J.~Noronha, and M.~Strickland, ``{New
  Exact Solution of the Relativistic Boltzmann Equation and its Hydrodynamic
  Limit},'' \href{http://dx.doi.org/10.1103/PhysRevLett.113.202301}{{\em Phys.
  Rev. Lett.} {\bf 113} (2014) no.~20, 202301},
\href{http://arxiv.org/abs/1408.5646}{{\tt arXiv:1408.5646 [hep-ph]}}.

\bibitem{Nopoush:2014qba}
M.~Nopoush, R.~Ryblewski, and M.~Strickland, ``{Anisotropic hydrodynamics for
  conformal Gubser flow},''
  \href{http://dx.doi.org/10.1103/PhysRevD.91.045007}{{\em Phys. Rev.} {\bf
  D91} (2015) no.~4, 045007},
\href{http://arxiv.org/abs/1410.6790}{{\tt arXiv:1410.6790 [nucl-th]}}.

\bibitem{Florkowski:2015lra}
W.~Florkowski, A.~Jaiswal, E.~Maksymiuk, R.~Ryblewski, and M.~Strickland,
  ``{Relativistic quantum transport coefficients for second-order viscous
  hydrodynamics},'' \href{http://dx.doi.org/10.1103/PhysRevC.91.054907}{{\em
  Phys. Rev.} {\bf C91} (2015)  054907},
\href{http://arxiv.org/abs/1503.03226}{{\tt arXiv:1503.03226 [nucl-th]}}.

\bibitem{Molnar:2016gwq}
E.~Molnar, H.~Niemi, and D.~H. Rischke, ``{Closing the equations of motion of
  anisotropic fluid dynamics by a judicious choice of a moment of the Boltzmann
  equation},'' \href{http://dx.doi.org/10.1103/PhysRevD.94.125003}{{\em Phys.
  Rev.} {\bf D94} (2016) no.~12, 125003},
\href{http://arxiv.org/abs/1606.09019}{{\tt arXiv:1606.09019 [nucl-th]}}.

\bibitem{Martinez:2017ibh}
M.~Martinez, M.~McNelis, and U.~Heinz, ``{Anisotropic fluid dynamics for Gubser
  flow},'' \href{http://dx.doi.org/10.1103/PhysRevC.95.054907}{{\em Phys. Rev.}
  {\bf C95} (2017) no.~5, 054907},
\href{http://arxiv.org/abs/1703.10955}{{\tt arXiv:1703.10955 [nucl-th]}}.

\bibitem{Damodaran:2017ior}
M.~Damodaran, D.~Molnar, G.~G. Barnafoldi, D.~Berenyi, and M.~Ferenc Nagy-Egri,
  ``{Testing and improving shear viscous phase space correction models},''
\href{http://arxiv.org/abs/1707.00793}{{\tt arXiv:1707.00793 [nucl-th]}}.

\bibitem{Romatschke:2009im}
P.~Romatschke, ``{New Developments in Relativistic Viscous Hydrodynamics},''
  \href{http://dx.doi.org/10.1142/S0218301310014613}{{\em Int. J. Mod. Phys.}
  {\bf E19} (2010)  1--53},
\href{http://arxiv.org/abs/0902.3663}{{\tt arXiv:0902.3663 [hep-ph]}}.

\bibitem{Ollitrault:2012cm}
J.-Y. Ollitrault and F.~G. Gardim, ``{Hydro overview},''
  \href{http://dx.doi.org/10.1016/j.nuclphysa.2013.01.047}{{\em Nucl. Phys.}
  {\bf A904-905} (2013)  75c--82c},
\href{http://arxiv.org/abs/1210.8345}{{\tt arXiv:1210.8345 [nucl-th]}}.

\bibitem{Gale:2013da}
C.~Gale, S.~Jeon, and B.~Schenke, ``{Hydrodynamic Modeling of Heavy-Ion
  Collisions},'' \href{http://dx.doi.org/10.1142/S0217751X13400113}{{\em Int.
  J. Mod. Phys.} {\bf A28} (2013)  1340011},
\href{http://arxiv.org/abs/1301.5893}{{\tt arXiv:1301.5893 [nucl-th]}}.

\bibitem{Jeon:2015dfa}
S.~Jeon and U.~Heinz, ``{Introduction to Hydrodynamics},''
  \href{http://dx.doi.org/10.1142/S0218301315300106}{{\em Int. J. Mod. Phys.}
  {\bf E24} (2015) no.~10, 1530010},
\href{http://arxiv.org/abs/1503.03931}{{\tt arXiv:1503.03931 [hep-ph]}}.

\bibitem{Jaiswal:2016hex}
A.~Jaiswal and V.~Roy, ``{Relativistic hydrodynamics in heavy-ion collisions:
  general aspects and recent developments},''
  \href{http://dx.doi.org/10.1155/2016/9623034}{{\em Adv. High Energy Phys.}
  {\bf 2016} (2016)  9623034},
\href{http://arxiv.org/abs/1605.08694}{{\tt arXiv:1605.08694 [nucl-th]}}.

\bibitem{Noronha:2015jia}
J.~Noronha and G.~S. Denicol, ``{Perfect fluidity of a dissipative system:
  Analytical solution for the Boltzmann equation in $\mathrm{AdS}_{2}\otimes
  \mathrm{S}_{2}$},'' \href{http://dx.doi.org/10.1103/PhysRevD.92.114032}{{\em
  Phys. Rev.} {\bf D92} (2015) no.~11, 114032},
\href{http://arxiv.org/abs/1502.05892}{{\tt arXiv:1502.05892 [hep-ph]}}.

\bibitem{Heller:2016rtz}
M.~P. Heller, A.~Kurkela, and M.~Spalinski, ``{Hydrodynamization and transient
  modes of expanding plasma in kinetic theory},''
\href{http://arxiv.org/abs/1609.04803}{{\tt arXiv:1609.04803 [nucl-th]}}.

\bibitem{Denicol:2016bjh}
G.~S. Denicol and J.~Noronha, ``{Divergence of the Chapman-Enskog expansion in
  relativistic kinetic theory},''
\href{http://arxiv.org/abs/1608.07869}{{\tt arXiv:1608.07869 [nucl-th]}}.

\bibitem{Bemfica:2017wps}
F.~S. Bemfica, M.~M. Disconzi, and J.~Noronha, ``{Causality and existence of
  solutions of relativistic viscous fluid dynamics with gravity},''
\href{http://arxiv.org/abs/1708.06255}{{\tt arXiv:1708.06255 [gr-qc]}}.

\bibitem{Florkowski:2017olj}
W.~Florkowski, M.~P. Heller, and M.~Spalinski, ``{New theories of relativistic
  hydrodynamics in the LHC era},''
\href{http://arxiv.org/abs/1707.02282}{{\tt arXiv:1707.02282 [hep-ph]}}.

\bibitem{Florkowski:2017jnz}
W.~Florkowski, E.~Maksymiuk, and R.~Ryblewski, ``{Coupled kinetic equations for
  quarks and gluons in the relaxation time approximation},''
\href{http://arxiv.org/abs/1710.07095}{{\tt arXiv:1710.07095 [hep-ph]}}.

\bibitem{Bjorken:1982qr}
J.~D. Bjorken, ``{Highly Relativistic Nucleus-Nucleus Collisions: The Central
  Rapidity Region},''
\href{http://dx.doi.org/10.1103/PhysRevD.27.140}{{\em Phys. Rev.} {\bf D27}
  (1983)  140--151}.

\bibitem{Alqahtani:2017jwl}
M.~Alqahtani, M.~Nopoush, R.~Ryblewski, and M.~Strickland, ``{3+1d
  quasiparticle anisotropic hydrodynamics for ultrarelativistic heavy-ion
  collisions},''
\href{http://arxiv.org/abs/1703.05808}{{\tt arXiv:1703.05808 [nucl-th]}}.

\bibitem{Bhatnagar:1954zz}
P.~L. Bhatnagar, E.~P. Gross, and M.~Krook, ``{A Model for Collision Processes
  in Gases. 1. Small Amplitude Processes in Charged and Neutral One-Component
  Systems},''
\href{http://dx.doi.org/10.1103/PhysRev.94.511}{{\em Phys. Rev.} {\bf 94}
  (1954)  511--525}.

\bibitem{Anderson:1974a}
J.~L. Anderson and H.~R. Witting, ``{A relativistic relaxation-time model for
  the Boltzmann equation},'' {\em Physica} {\bf 74} (1974)  466.

\bibitem{Anderson:1974b}
J.~L. Anderson and H.~R. Witting, ``{Relativistic quantum transport
  coefficients},'' {\em Physica} {\bf 74} (1974)  489.

\bibitem{Czyz:1986mr}
W.~Czyz and W.~Florkowski, ``{Kinetic Coefficients for Quark - Anti-quark
  Plasma},''
{\em Acta Phys. Polon.} {\bf B17} (1986)  819--837.

\bibitem{LLfluid}
L.~D. Landau and E.~M. Lifshitz, {\em Fluid Mechanics, Second Edition: Volume 6
  (Course of Theoretical Physics)}.
\newblock 1987.

\bibitem{Florkowski:2010cf}
W.~Florkowski and R.~Ryblewski, ``{Highly-anisotropic and strongly-dissipative
  hydrodynamics for early stages of relativistic heavy-ion collisions},''
  \href{http://dx.doi.org/10.1103/PhysRevC.83.034907}{{\em Phys. Rev.} {\bf
  C83} (2011)  034907},
\href{http://arxiv.org/abs/1007.0130}{{\tt arXiv:1007.0130 [nucl-th]}}.

\bibitem{Martinez:2010sc}
M.~Martinez and M.~Strickland, ``{Dissipative Dynamics of Highly Anisotropic
  Systems},'' \href{http://dx.doi.org/10.1016/j.nuclphysa.2010.08.011}{{\em
  Nucl. Phys.} {\bf A848} (2010)  183--197},
\href{http://arxiv.org/abs/1007.0889}{{\tt arXiv:1007.0889 [nucl-th]}}.

\bibitem{Romatschke:2003ms}
P.~Romatschke and M.~Strickland, ``{Collective modes of an anisotropic quark
  gluon plasma},'' \href{http://dx.doi.org/10.1103/PhysRevD.68.036004}{{\em
  Phys. Rev.} {\bf D68} (2003)  036004},
\href{http://arxiv.org/abs/hep-ph/0304092}{{\tt arXiv:hep-ph/0304092
  [hep-ph]}}.

\bibitem{Florkowski:2011jg}
W.~Florkowski and R.~Ryblewski, ``{Projection method for boost-invariant and
  cylindrically symmetric dissipative hydrodynamics},''
  \href{http://dx.doi.org/10.1103/PhysRevC.85.044902}{{\em Phys. Rev.} {\bf
  C85} (2012)  044902},
\href{http://arxiv.org/abs/1111.5997}{{\tt arXiv:1111.5997 [nucl-th]}}.

\bibitem{Florkowski:2008ag}
W.~Florkowski, ``{Anisotropic fluid dynamics in the early stage of relativistic
  heavy-ion collisions},''
  \href{http://dx.doi.org/10.1016/j.physletb.2008.07.101}{{\em Phys. Lett.}
  {\bf B668} (2008)  32--35},
\href{http://arxiv.org/abs/0806.2268}{{\tt arXiv:0806.2268 [nucl-th]}}.

\bibitem{Bialas:1984wv}
A.~Bialas and W.~Czyz, ``{Boost Invariant Boltzmann-vlasov Equations for
  Relativistic Quark - Anti-quark Plasma},''
\href{http://dx.doi.org/10.1103/PhysRevD.30.2371}{{\em Phys. Rev.} {\bf D30}
  (1984)  2371}.

\bibitem{Bialas:1987en}
A.~Bialas, W.~Czyz, A.~Dyrek, and W.~Florkowski, ``{Oscillations of Quark -
  Gluon Plasma Generated in Strong Color Fields},''
\href{http://dx.doi.org/10.1016/0550-3213(88)90035-1}{{\em Nucl. Phys.} {\bf
  B296} (1988)  611--624}.

\bibitem{Nopoush:2014pfa}
M.~Nopoush, R.~Ryblewski, and M.~Strickland, ``{Bulk viscous evolution within
  anisotropic hydrodynamics},''
  \href{http://dx.doi.org/10.1103/PhysRevC.90.014908}{{\em Phys. Rev.} {\bf
  C90} (2014) no.~1, 014908},
\href{http://arxiv.org/abs/1405.1355}{{\tt arXiv:1405.1355 [hep-ph]}}.

\end{thebibliography}\endgroup
\bibliographystyle{utphys}

\newpage

\end{document}